\documentclass[showpacs, twocolumn]{revtex4-1}
\usepackage{graphicx}
\usepackage{amsmath}
\usepackage{appendix}

\begin{document}

\title{ Bose-Bose mixtures in a weak disorder potential: Fluctuations and Superfluidity}

\author{Abdel\^{a}ali Boudjem\^{a}a$^{1,2}$ and Karima Abbas$^{1,2}$}
\affiliation{$^1$Department of Physics, Faculty of Exact Sciences and Informatics, Hassiba Benbouali University of Chlef, P.O. Box 78, 02000, Ouled-Fares, Chlef, Algeria. \\
$^2$Laboratory of Mechanics and Energy, Hassiba Benbouali University of Chlef, P.O. Box 78, 02000, Ouled-Fares, Chlef, Algeria.}
\email {a.boudjemaa@univ-chlef.dz}


\begin{abstract}

We study the properties of a homogeneous dilute Bose-Bose gas in a weak disorder potential at zero temperature. 
By using the perturbation theory, we calculate the disorder corrections to the condensate density, the equation of state, the compressibility, and the superfluid density 
as a function of density, strength of disorder, and miscibility parameter. 
It is found that the disorder potential may lead to modifying the miscibility-immiscibility condition and  
a full miscible phase turns out to be impossible in the presence of the disorder.
We show that the intriguing interplay of the disorder and intra- and interspecies interactions may strongly influence the localization of each component, 
the quantum fluctuations, and the compressibility, as well as the superfluidity of the system.

\end{abstract}

\maketitle

\section{Introduction}

In recent years, degenerate multi-component quantum gases have prompted considerable interest in the community of cold atoms physics 
both theoretically and experimentally due to their rich phase diagram. 
One of the most significant characteristics of such multi-component structures is their miscibility-immiscibility transition which depends on 
the ratio of the intra- and interspecies interactions \cite{Tin, Pu, Papp}, on the condensate numbers \cite{Lee}, and on thermal fluctuations \cite{Lell,Roy, Boudj2018, Ota}.
A mixture of two-component Bose-Einstein condensate (BEC) plays a crucial role in various systems, such as solitons (see e.g. \cite{Kev}),
vortices (see e.g. \cite{Matt}), and bilayer Bose systems (see e.g. \cite {Wang, Boudj20}). 
Very recently, it has been found that the balance between the mean-field term and the beyond- mean-field quantum fluctuation
may lead to the formation of a mixture droplet phase \cite{Petrov,Cab, Sem, Boudj18}. 

On the other hand, the creation of disorder using speckle lasers \cite {Lye, Billy} or incommensurate laser beams \cite{Dam, Roat} 
opens promising new avenues in condensed matter physics and in the ultracold quantum gases field.
The competition between disorder and interactions plays a nontrivial role in developing a fundamental understanding of many aspects
of ultracold gases namely: the Bose glass (a gapless compressible insulating state) \cite{Ma, Giam, Fich, Scal, Krauth}, 
Anderson localization \cite {Lye, Billy,Dam, Roat, Schut, Laur, Lugan1,Ski,Lugan,Gaul}, 
disordered BEC in optical lattices \cite {Wh, Pas, Deis, Alei},  Bose-Fermi mixtures \cite {Ahuf,Franç}, and dipolar BEC in random potentials 
\cite{Boudj2018, Krum, Nik,Ghab, Boudj,Boudj1,Boudj2,Boudj3,Boudj4}. 

Until now, there has been little work treating disordered ultracold Bose-Bose mixtures.  
A general mechanism of random-field-induced order has been analyzed in both lattice \cite{Wehr} and continuum \cite{Nied}  two-component BEC. 
Localization of a trapped two-component BEC in a one-dimensional random potential has been numerically addressed in Ref.\cite {Xi}.
It has been found in addition that disorder plays a crucial role in the dynamics of spin-orbit coupled BEC in a random potential \cite{Mard}.

This paper aims to investigate the impacts of a weak disorder potential on the quantum fluctuations and on the superfluidity of two-component BEC.
To this end, we extend the perturbative theory applicable to the single component bosonic gas \cite{Lugan, Gaul, Krum, Nik,Boudj4,LSP} and present a detailed 
analysis of weakly interacting homogeneous two-component Bose gases subjected to weak disorder potential with delta-correlation function.
The effects of the disorder on the miscibility-immiscibility condition are also deeply investigated. 
This study not only bridges the gap between superfluidity, interactions and disorder but also it is
important from the viewpoint of elucidating the localization phenomenon of two bosonic species. 

We derive useful expressions for the condensate fluctuations due to the disorder known as {\it glassy fraction}, the equation of state (EoS), the compressibility, and the superfluid density. 
We look at how each species is influenced by the disorder and how the interaction between disordered bosons influences the coupling and the phase transition between the two components.
Our results reveal that the localization of each species does not depend only on the disorder strength but 
depends also on the interspecies interactions and the ratio of intraspecies interactions. 
We show that the disorder effects could significantly enhance chemical potential of each species.
The disorder corrections to the superfluid density show a similar behavior as the glassy fraction of the condensate.
Moreover, we obtain disorder corrections to the compressibility, and the miscibility condition and accurately determine the critical disorder strength 
above which a transition from miscible to immiscible phase occurs.
In the decoupling regime where the interspecies interaction goes to zero, we find good agreement with the analytical results obtained 
within the Huang-Meng-Bogoliubov model \cite{HM} and perturbative theory for a single component BEC. 
Experimental evidence of the Huang-Meng theory for a single BEC has been reported most recently in Ref.\cite{Nagler}.

The rest of this paper is structured as follows. 
In Sec.\ref{flism}  we develop the perturbative theoretical description with respect to disorder which is based on the coupled Gross-Piteavskii (GP) equations and discuss its validity.
Section \ref{GFF} deals with the fluctuations due to the disorder potential.
We focus explicitly on the effects of weak delta-function correlated disorder and derive an analytical formula for the glassy fraction.
Its behavior is deeply highlighted as a function of the miscibility parameter and interspecies interactions.
In Sec.\ref{EoS} we calculate the disorder corrections to the EoS by extending the renormalization scheme used in a dirty single BEC \cite {Nik,Boudj4}.
Section \ref{MC} is dedicated to investigating the compressibility and to establishing the miscibility condition for a disordered homogeneous mixture. 
We find that a binary Bose miscible mixture cannot occur in the presence of the disorder.
In Sec.\ref{SF} we look at how a weak disorder potential influences the superfluidity. 
Section\ref {Conc} contains some conclusions and outlooks.

\section{Model} \label{flism}

Consider weakly interacting binary Bose gases in a weak random potential fulfilling mean-field miscibility criterion (see below).
The system is described by the coupled GP equations \cite{Roy, Boudj2018, Boudj18, Boudj5}
\begin{align} 
\mu_j \Phi_j = \left[- \frac{\displaystyle\hbar^2}{2m_j}\nabla^2 +U_j+g_j |\Phi_j |^2 + g_{12}  |{\Phi}_{\overline j} |^2 \right]\Phi_j,  \label{GPE}
\end{align}
where $\Phi_j$ is the wavefunction of each condensate, the indice $j$ is the species label, $\overline{j}=3-j$,  $\mu_j$ is the chemical potential of each condensate, 
$g_j=(4\pi \hbar^2/m_j) a_j$ and $g_{12}=g_{21}= 2\pi \hbar^2 (m_1^{-1}+m_2^{-1}) a_{12}$ with 
$a_j$ and $a_{12}$ being the intraspecies and the interspecies scattering lengths, respectively.
The gas parameter satisfies the condition $n_ja_j^3\ll1$.
The disorder potential $U_j({\bf r})$ is described by vanishing ensemble averages $\langle U(\mathbf r)\rangle=0$, and a finite correlation of the form
$\langle U(\mathbf r) U(\mathbf r')\rangle=R (\mathbf r-\mathbf r')$.

For weak disorder, Eq.(\ref{GPE}) can be solved using straightforward perturbation theory
in powers of $U$ using the expansion \cite{Lugan, Gaul, Krum, Nik,Boudj4,LSP} 
\begin{equation}  \label{Exps}
\Phi_j=\Phi_j^{(0)}+\Phi_j^{(1)} ({\bf r})+\Phi_j^{(2)} ({\bf r})+\cdots, \;\;\;\;\   j=1,2 
\end{equation}
where the index $i$ in the real valued functions $\Phi^{(i)}({\bf r})$ signals the $i$-th order contribution with respect to the disorder potential. 
They can be determined by inserting the perturbation series (\ref{Exps}) into Eq.(\ref{GPE}) and by collecting the terms up to $U^2$.
The zeroth order gives 
\begin{equation} \label{GPE0} 
\Phi_j^{(0)}=  \sqrt{\frac {\mu_j - g_{12} \Phi_{\overline j}^{(0)2} } { g_j}}, 
\end{equation} 
which is the uniform solution in the absence of a disorder potential. 
Combining Eqs.(\ref{GPE0}), yields
\begin{equation} \label{GPE01} 
\Phi_j^{(0)}=  \sqrt{\frac {\mu_j} { g_j} \left(1-\frac{g_{12}} {g_{\overline j}} \frac{\mu_{\overline j}} {\mu_j} \right) \frac{\Delta}{\Delta-1}}, 
\end{equation} 
where $\Delta=g_j g_{\overline j}/g_{12}^2$  is the miscibility parameter which characterizes the miscible-immiscible transition.
For  $\Delta >1$, the mixture is miscible while it is immiscible for $\Delta <1$.\\
The first-order equation reads 
\begin{align}  \label{GPE1}
&-\frac{\hbar^2}{2m_j} \nabla^2 \Phi_j^{(1)} ({\bf r})+U_j({\bf r}) \Phi_j^{(0)}+ 2 g_j \Phi_j^{(0)2} \Phi_j^{(1)} ({\bf r})\\
&+2g_{12} \Phi_j^{(0)}\Phi_{\overline j}^{(0)}\Phi_{\overline j}^{(1)}({\bf r}) =0  \nonumber,
\end{align}
Performing a Fourier transformation, one obtains
\begin{equation}   \label{GPE11}
\Phi_j^{(1)}({\bf k})= - \frac{ \left[U_j({\bf k})+2g_{12} \Phi_{\overline j}^{(0)} \Phi_{\overline j}^{(1)}({\bf k}) \right] \Phi_j^{(0)}} {E_{kj}+2 g_j\Phi_j^{(0)2} }, 
\end{equation}
where $E_{kj}=\hbar^2k^2/2m_j$. \\ 
For $E_{kj} \ll 2 g_j\Phi_j^{(0)2}=\mu_j \left(1-g_{12} \mu_{\overline j} /g_{\overline j} \mu_j \right) \Delta/(\Delta-1)$, 
the kinetic energy is negligible compared to the random potential energy then, the mixture deformation sustains only the potential effects.
Therefore, the coupled GP equations  (\ref{GPE}) yield for the total density
$n_j ({\bf r})= \Phi^{(0)2}_j+n_j^{(1)} ({\bf r})$, where $n_j^{(1)}= \Delta \big ( n_{0j}-  g_{12}n_{0 \overline j} /g_j \big)/(\Delta -1)$,
with $n_{0j}=(\mu_j-V_j)/g_j$ being the decoupled condensate density which is nothing else than the standard Thomas-Fermi-like shape.
For $E_k \gg \mu_j \left(1-g_{12} \mu_{\overline j} /g_{\overline j} \mu_j \right) \Delta/(\Delta-1)$, 
the densities of the two BEC follow the modulations of a smoothed  disorder potential where  the variations of $U$ have been smoothed out.

The second-order term is governed by the following equation
\begin{align}  \label{GPE2}
&-\frac{\hbar^2}{2m_j} \nabla^2 \Phi_j^{(2)} ({\bf r})+U_j({\bf r}) \Phi_j^{(1)}+  g_j\bigg [2\Phi_j^{(0)2} \Phi_j^{(2)} ({\bf r}) \\
&+3\Phi_j^{(0)} \Phi_j^{(1)2} ({\bf r})\bigg]+g_{12} \bigg[2\Phi_{\overline j}^{(0)}\Phi_{\overline j}^{(1)}({\bf r})  \Phi_j^{(1)} ({\bf r})+  \Phi_j^{(0)} \Phi_{\overline j}^{(1)2} ({\bf r})   \nonumber \\
&+ 2 \Phi_j^{(0)} \Phi_{\overline j}^{(0)} \Phi_{\overline j}^{(2)}  ({\bf r}) \bigg]=0.\nonumber 
\end{align}
The solution of this equation in the momentum space reads
\begin{align} \label{GPE02}
\Phi_j^{(2)}({\bf k})&= - \int \frac{d \bf k'}{(2\pi)^3}   \frac{\Phi_j^{(1)} ({\bf k- k'}) \left[ U_j({\bf k'}) +3g_j \Phi_j^{(0)} \Phi_j^{(1)} ({\bf k'}) \right] } {E_{kj}+2 g_j\Phi_j^{(0)2} }  
\nonumber \\
&- g_{12}\frac {2\Phi_{\overline j}^{(0)} \Phi_j^{(0)}  \Phi_{\overline j}^{(2)}({\bf k}) }  {E_{kj}+2 g_j\Phi_j^{(0)2} }- g_{12}\int \frac{d \bf k'}{(2\pi)^3}  \Phi_{\overline j}^{(1)} ({\bf k- k'})  \nonumber\\
& \times \frac{\left[ 2\Phi_{\overline j}^{(0)} \Phi_j^{(1)}({\bf k'}) + \Phi_{\overline j}^{(1)} ({\bf k'}) \Phi_j^{(0)}\right ] } {E_{kj}+2 g_j\Phi_j^{(0)2} }. 
\end{align} 
Equation (\ref{GPE02}) enables us to selfconsistently determine the chemical potential of the system (see below).

Finally, the validity of the present perturbation approach requires  the condition: 
$U \ll g_j \Phi_j^{(0)2} \simeq g_j n_j$, where $\Phi_j^{(0)}$ is given in Eq.(\ref{GPE01}),
tells us that the densities do not vary much around the homogeneous values.
For $g_{12}=0$, one recovers the well-known condition ($U\ll g\Phi^{(0)2}$) established for a disordered single BEC \cite{LSP}.
Indeed, this simple assumption indicates how localization can be destroyed in a regime of weak interactions.
However, the perturbation approach is no longer valid in the regime of strong disorder.


\section {Glassy fraction} \label{GFF}

In this section we deal with the mixture fluctuations due to the disorder potential.
It has been shown  that the disorder contribution to the condensate can be given as the variance of the wavefunction
$n_{Rj}= n_j-n_{cj}$ \cite{Krum,Nik}, where 
\begin{equation}  \label{totD}
n_j= \langle \Phi_j ^2({\bf r}) \rangle= \Phi_j^{(0)2}+ \langle \Phi_j^{(1)2} ({\bf r}) \rangle+2\Phi_j^{(0)}  \langle \Phi_j^{(2)} ({\bf r}) \rangle +\cdots
\end{equation}
and
\begin{equation} \label{conD}
n_{cj}= \langle \Phi_j ({\bf r}) \rangle^2= \Phi_j^{(0)2}+ 2\Phi_j^{(0)}\langle \Phi_j^{(2)} ({\bf r}) \rangle +\cdots
\end{equation}
is the condensed density.
Subtracting  (\ref{conD}) from (\ref{totD}), one obtains $n_{Rj}=\langle \Phi_j^{(1)2} ({\bf r})  \rangle+\cdots$, which is 
in fact analog to the Edwards-Anderson order parameter of a spin glass  \cite{Yuk, Nik, Edw}.

From now on, we shall consider $U_1=U_2=U$ and $m_1=m_2=m$.

Employing the Fourier transform of $\Phi_j^{(1)} ({\bf r})$ i.e. Eq. (\ref{GPE11}), and using the fact that $\langle U({\bf k'})U({\bf k''}) \rangle=(2\pi)^{3}R({\bf k'})\delta({\bf k'+k''})$,
the glassy fraction, $n_{Rj}$, can be written as:
\begin{align}\label{GF}
&n_{Rj}= n_j\int\frac{\bf dk}{(2\pi)^3} R({\bf k}) \left [ \frac {E_k+2n_{\overline j} \big(g_{\overline j}-g_{12} \big)}  { {\cal E}_k } \right]^2, 
\end{align}
where ${\cal E}_k=\big(E_k+2g_j n_j \big) \big(E_k+2g_{\overline j} n_{\overline j}\big) - 4g_{12}^2 n_j n_{\overline j}$.

For analytical tractability, we consider the white noise random potential, which assumes a delta distribution  
\begin{equation}  \label{delt}
R ({\bf r}- {\bf r'})=R_0\delta ({\bf r}- {\bf r'}), 
\end{equation}
where $R_0$ is the disorder strength with dimension (energy)$^2$ $\times$ (length)$^3$. 
The model (\ref{delt}) is valid when the correlation length of the correlation function $R ({\bf r}- {\bf r'})$ is sufficiently shorter than the healing length. \\
After some algebra, we get a useful formula for the glassy fraction: 
\begin{equation}\label{GF1}
\frac{n_{Rj}}{n_j}= 4\pi R_j' \sqrt{\frac{ n_j a_j^3}{\pi}}  f_j(\Delta),
\end{equation}
where $R_j'=R_0/g_j^2n_j$ is a dimensionless disorder strength and, 
\begin{align}\label{disf}
f_j(\Delta)&= \left[ \frac{(2\beta_j)^{-3/2}}{\sqrt{1+\bar\mu_j+\sqrt{\beta_j}}} -\frac{(2\beta_j)^{-3/2}}{\sqrt{(1+\bar\mu_j)-\sqrt{\beta_j}}} \right] \bar f_1(\Delta) \\
&+ \left[ \frac{\sqrt{2}\beta_j^{-1}}{4\sqrt{1+\bar\mu_j+\sqrt{\beta_j}}}+\frac{\sqrt{2}\beta_j^{-1}}{4\sqrt{(1+\bar\mu_j)-\sqrt{\beta_j}}} \right] \bar f_2(\Delta) \nonumber
\end{align}
where \\
$\bar f_1(\Delta)=(1+\bar\mu_j)^{3}+2\alpha_j (1+\bar\mu_j)^2-4(1+\bar\mu_j)\big [2\bar\mu_j\left(\frac{\Delta-1}{\Delta}\right)+\alpha_j^{2}\big]
+8\bar\mu_j\alpha_j\left(\frac{\Delta-1}{\Delta}\right)$, \\
$\bar f_2(\Delta)=(1+\bar\mu_j)^{2}+2\alpha_j (1+\bar\mu_j)-6\bar\mu_j\big(\frac{\Delta-1}{\Delta}\big)-2\alpha_j^{2}$, \\
$\beta_j=(1+\bar\mu_j)^{2}-4\bar\mu_j\left[(\Delta-1)/\Delta\right]$, \\
$\alpha_j=\bar\mu_j \left(1-\sqrt{g_{j}/\big(g_{\overline j}\Delta \big) }\right)$, and $\bar\mu_j= n_{\overline j} g_{\overline j}/ n_jg_j$. \\
Equation (\ref{GF1}) is appealing since it describes the glassy fraction in terms of the miscibility parameter.
The total disorder density is given by $n_R= n_{R1}+n_{R2}$.
For $\Delta \rightarrow  \infty$ (or $g_{12} \rightarrow 0$, equivalently), we find from Eq.(\ref{GF1}) that $f_1(\infty)=f_2(\infty)= 1/2$.  
Therefore, we should reproduce the famous Huang and Meng result \cite{HM}, $n_R/n= 2\pi R' \sqrt{ n a^3/\pi}$ for the single component disorder fraction. 
The intriguing interplay between the strong intercomponent coupling and the disorder effects in the regime $\Delta-1 \ll 1$ would cause a sharp increase in the functions $f_j(\Delta)$.
Near the phase separation i.e. $\Delta \rightarrow 1$ (or $g_{12} \rightarrow \sqrt{g_1g_2}$, equivalently), the functions $f_j(\Delta)$ are diverging. 
They are complex for $\Delta <1$ and hence, the mixture undergoes instability.

\begin{figure}[htb] 
\includegraphics[scale=0.8]{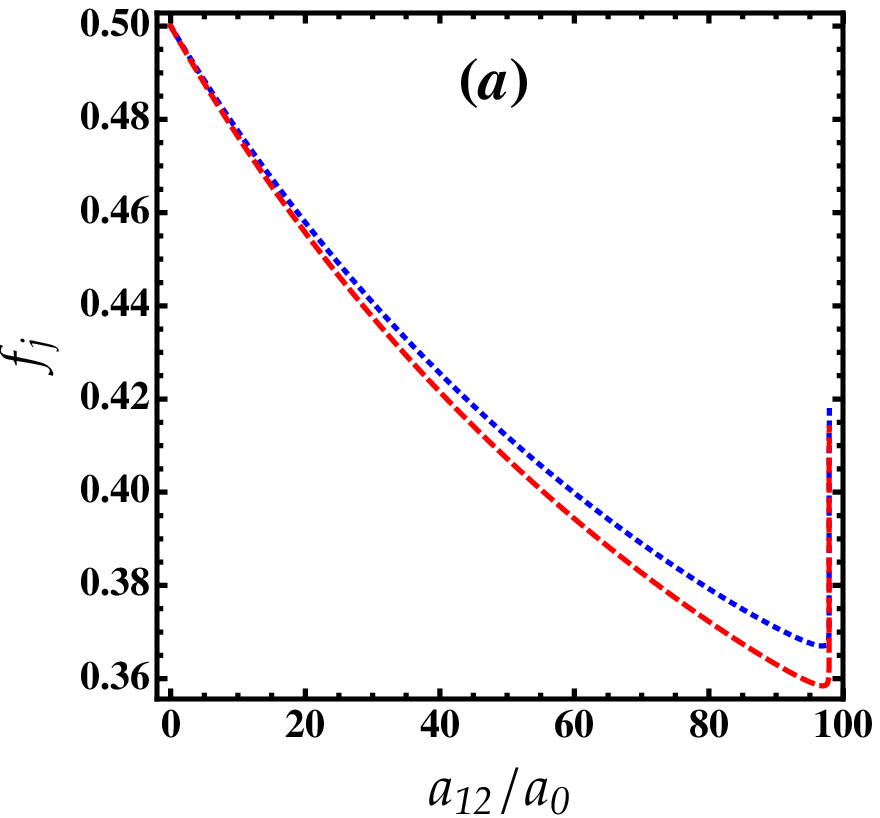}
\includegraphics[scale=0.8]{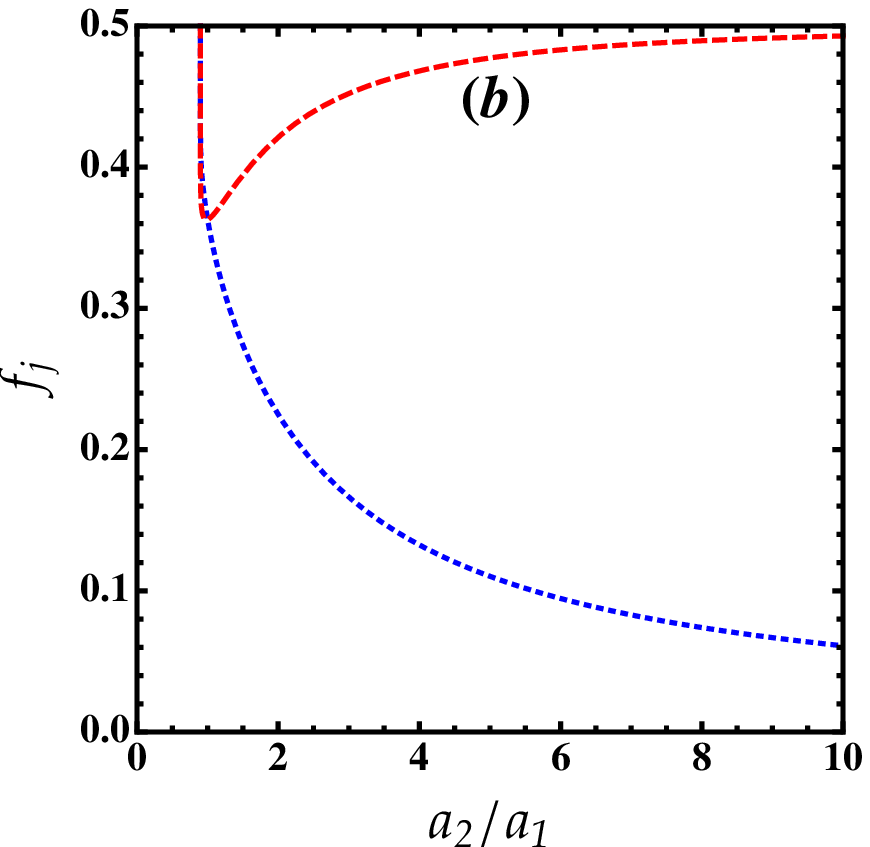}
\caption {(Color online)  (a) Behavior of the disorder functions $f_j$ as a function of the interspecies interaction strength $a_{12}$ for ${}^{87}$Rb-${}^{87}$Rb mixture. 
(b) Behavior of the disorder functions $f_j$ as a function of the ratio $a_2/a_1$ for $a_{12}=90 a_0$. 
Blue dotted lines: $f_1$. Red dashed lines: $f_2$.
Here $a_{12}$ can be adjusted via Feshbach resonance.}
\label{dis}
\end{figure}

The disorder functions $f_j$ have the following asymptotic behavior for small $a_{12}$
$$f_j(a_{12})= \frac{1}{2}-\frac{n_{\overline j}} {n_j a_j \left(1+\sqrt{ \frac{n_{\overline j} a_{\overline j}}{n_j a_j}}\right)^2}a_{12}+ \cdots,$$
and for large $a_{12}$
$$f_j(a_{12})= \frac{\left(\sqrt{a_{\overline j}/a_j}-1\right)^2} {2\left(\frac{n_{\overline j} a_{\overline j}}{n_j a_j}+1\right)^{3/2} \sqrt{\left(a_j a_{\overline j}/a_{12}^2\right)-1}}+ \cdots.$$
It is straightforward to check that these asymptotic results perfectly agree with the solutions shown in Fig. \ref{dis} (a) in the asymptotic regime. 

As an illustration of our theoretical formalism, we consider a two-component Bose condensate of rubidium atoms in two different internal states ${}^{87}$Rb-${}^{87}$Rb.
We have taken the intra-component scattering lengths : $a_1= 100.4\,a_0$ and $a_2= 95.44\,a_0$ ($a_0$ is the Bohr radius) \cite{Ego},
and the densities:  $n_1=1.5 \times 10^{21}$ m$^{-3}$, and $n_2=10^{21}$ m$^{-3}$.  Thus, the parameter $n_ja_j^3$ is as small as $ \sim 10^{-4}$.

Figure \ref{dis} (a) shows that for $a_{12}/a_0 \leq 97.89$, the functions $f_j$ are decreasing  with the interspecies interaction giving rise to the delocalization of both species.
In the vicinity of the transition between the miscible and immiscible phases i.e. $a_{12}/a_0 = 97.89$, 
the functions $f_j$ exhibit an anomalous behavior where they develop a small minimum.
Then they start to increase for $a_{12} /a_0> 97.89$.  In such a regime, both species are srongly localized in the local wells of the random potential.

The situation is quite different for fixed $a_{12}$ and varying the interactions ratio $a_2/a_1$.
The disorder functions  $f_1$ and $f_2$ decrease/increase with the ratio $a_2/a_1$ as is shown in Fig.\ref{dis} (b).
The function $f_2$ develops a minimum at $a_2 \simeq a_1$.
For $a_2/a_1 \gtrsim 5$, $f_1$ is very small and thus, the first component becomes almost superfluid due to the suppression of the localization, while the second BEC remains localized
regardless of the value of $a_{12}$.
One can conclude that the localization of one component does not  trigger the localization of the second component
due to the interplay of the intra- and interspecies interactions and the disorder potential.

\section{Equation of state} \label{EoS}

The EoS can be calculated by substituting Eqs.(\ref{GPE0})-(\ref{GPE02}) into Eq.(\ref{totD}) and solving 
the equation $\langle \Phi_j^2 (\mu_{bj} )\rangle=n (\mu_{bj} )$, where $\mu_{bj}$ represents 
the bare chemical potential. It diverges for uncorrelated disorder \cite {Nik,Boudj4}. We then obtain
\begin{widetext} 
\begin{align} \label {eos1}
\mu_{bj} (n_{j},n_{\overline j})&=g_{j}n_{j}+g_{12}n_{\overline j}- \int\frac{d{\bf k}}{(2\pi)^{3}}\frac{R({\bf k})} {(g_{j}g_{\overline j}-g_{12}^{2}) {\cal E}_k} 
\Bigg\{ (g_j g_{\overline j}-g_{12}^2) [E_k-2n_{\overline j}(g_{12}-g_{\overline j})] +g_{12}g_j [E_k-2n_j (g_{12}-g_j)]\nonumber\\
&-\frac{2g_{12}g_j g_{\overline j} n_{\overline j} [E_k-2n_j (g_{12}-g_j)]^2-2g_j^2 g_{\overline j} n_j [E_k-2n_{\overline j} (g_{12}-g_{\overline j})]^2} {{\cal E}_k} 
-2g_{12}n_{\overline j}(g_j g_{\overline j}-g_{12}^2) [E_k-2n_j (g_{12}-g_j)] \nonumber\\
& \times \frac{[E_k-2n_{\overline j}(g_{12}-g_{\overline j})]} {{\cal E}_k} -\frac{2g_{12}g_j g_{\overline j} n_j^{3/2} [E_k-2n_{\overline j} (g_{12}-g_{\overline j})]^3} {(g_j g_{\overline j}-g_{12}^2) 
 {\cal E}_k^2}\Bigg\}.  
\end{align}
\end{widetext} 
To overcome this unphysical ultraviolet divergence, we renormalize the chemical potential. 
The renormalized chemical potential is defined as:
\begin{equation}\label{eos2}
\mu_{j}(n_{j},n_{\overline j})=\mu_{bj}(n_{j},n_{\overline j})-\mu_{bj}(0),
\end{equation}
where
\begin{equation}\label{eos3}
 \mu_{bj}(0)=-\int\frac{d \bf k}{(2\pi)^{3}} R({\bf k}) \left[\frac{1}{E_k}+\frac{g_j g_{12}}{(g_j g_{\overline j}-g_{12}^{2})E_k}\right].
\end{equation}
Omitting higher order in $g_{12}$, we obtain, in second-order of the disorder strength, the following renormalized EoS
\begin{align}
&\mu_j=g_j n_j+g_{12} n_{\overline j}+ \int\frac{d \bf k}{(2\pi)^{3}}\frac{R(k)}{ (g_j g_{\overline j}-g_{12}^{2}) {\cal E}_k^2 E_k} \\
&\times \Bigg\{ 4g_j^{2}g_{\overline j} n_j \left(E_k+g_j n_j\right) (E_k+2g_{\overline j} n_{\overline j})^{2} \nonumber\\
&+4g_j g_{\overline j} g_{12}n_{\overline j} \bigg[(E_k+g_{\overline j} n_{\overline j})(E_k+2g_j n_j)^2 \nonumber\\
&+E_k^{2} (E_k+2g_{\overline j} n_{\overline j} )\bigg]\bigg\} \nonumber.
\end{align}
This equation allows us to calculate the sound velocity and the inverse compressibility.

\begin{figure}[htb] 
\includegraphics[scale=0.8]{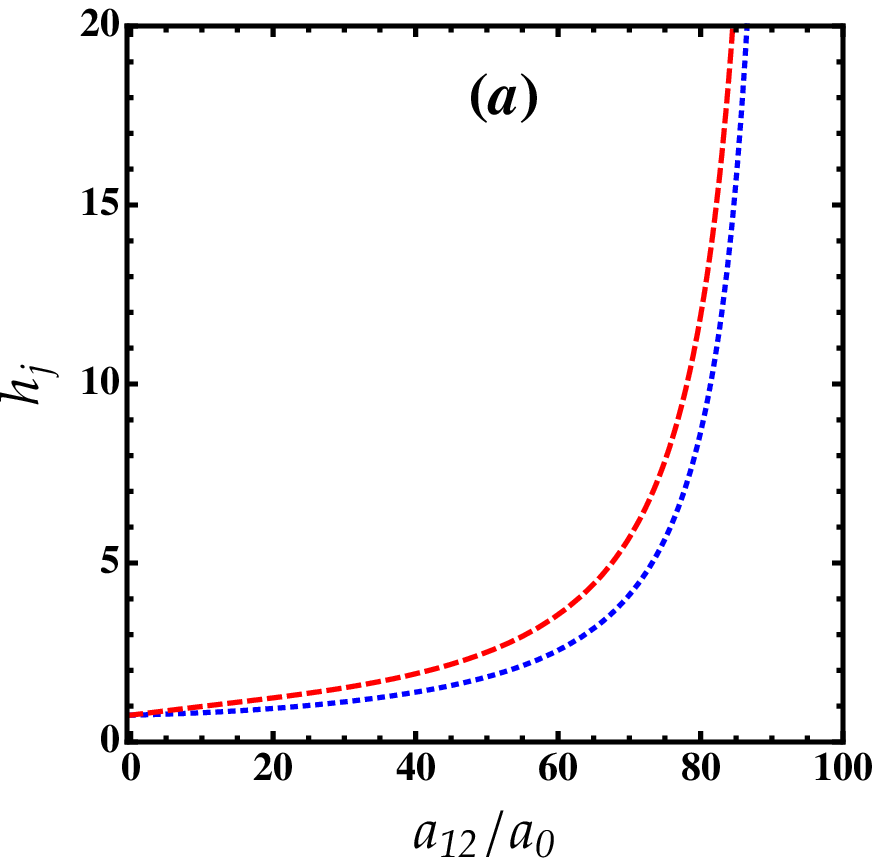}
\includegraphics[scale=0.81]{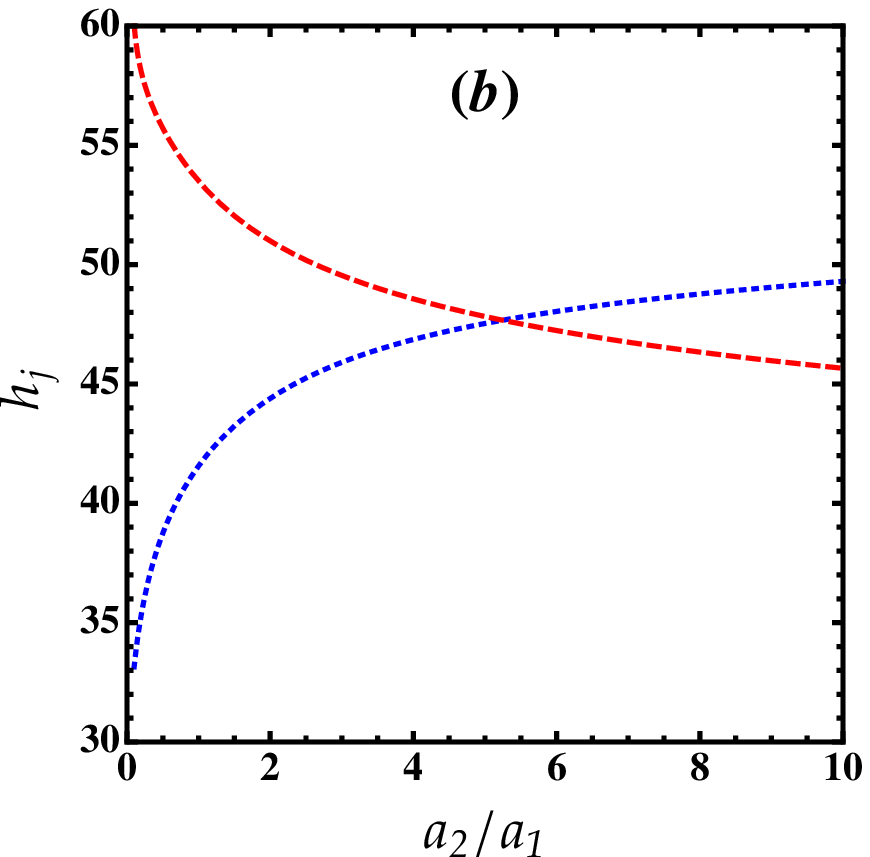}
\caption {(Color online) (a) Behavior of the disorder functions $h_j$ as a function of $a_{12}$ for ${}^{87}$Rb-${}^{87}$Rb mixture.
(b) Behavior of the disorder functions $h_j$ as a function of the ratio $a_2/a_1$ for $a_{12}=90 a_0$.
Blue dotted lines: $h_1$. Red dashed lines: $h_2$.}
\label{dis11}
\end{figure}

For delta-correlated disorder (\ref{delt}), the EoS reads
\begin{equation}\label{GF2}
\mu_j=g_j n_j+g_{12}n_{\overline j}+ 16\pi g_j n_j R_j' \sqrt{\frac{ n_j a_j^3}{\pi}} h_j(\Delta),
\end{equation}
where \\
\begin{equation}\label{GF3}
h_j(\Delta)=\frac{1}{(2\beta_j)^{3/2}} \frac{\Delta}{\Delta-1} \left[\bar h_1(\Delta)+\frac{n_{\overline j} g_{12}}{n_j g_j} \bar h_2(\Delta)\right],
\end{equation}
and 
\begin{align} 
\bar h_1(\Delta)&= \left(\frac{1}{\sqrt{1+\bar\mu_j+\sqrt{\beta_j}}}-\frac{1}{\sqrt{(1+\bar\mu_j)-\sqrt{\beta_j}}} \right) H_1(\Delta) \nonumber\\
&+\left( \frac{\sqrt{\beta_j} }{\sqrt{1+\bar\mu_j+\sqrt{\beta_j}}}+\frac{\sqrt{\beta_j} }{\sqrt{(1+\bar\mu_j)-\sqrt{\beta_j}}} \right) H_2(\Delta),\nonumber \\
\bar h_2(\Delta)&= \left( \frac{1}{\sqrt{1+\bar\mu_j+\sqrt{\beta_j}}}-\frac{1}{\sqrt{(1+\bar\mu_j)-\sqrt{\beta_j}}} \right) H_3(\Delta) \nonumber\\
&+\left( \frac{\sqrt{\beta_j} }{\sqrt{1+\bar\mu_j+\sqrt{\beta_j}}}+\frac{\sqrt{\beta_j} }{\sqrt{(1+\bar\mu_j)-\sqrt{\beta_j}}} \right) H_4 (\Delta), \nonumber
\end{align}
where\\
$H_1(\Delta)=(1+\bar\mu_j)^3+(1+\bar\mu_j)^2\left[\frac{1}{2}+2\bar\mu_j-\frac{1}{2}\bar\mu_j\left(\frac{\Delta}{\Delta-1}\right)\right]
-4(1+\bar\mu_j)\left[\bar\mu_j^2+\bar\mu_j\left(1+2\left(\frac{\Delta-1}{\Delta}\right)\right)\right] +2 \bar\mu_j\left(\frac{\Delta-1}{\Delta}\right)(1+4\bar\mu_j)+6\bar\mu_j^2$,\\
$H_2(\Delta)=(1+\bar\mu_j)^2+\frac{1}{2} (1+\bar\mu_j)\left[1+\bar\mu_j\left(\frac{\Delta}{\Delta-1}\right)\right]-6\bar\mu_j\left(\frac{\Delta-1}{\Delta}\right)$,\\
$H_3(\Delta)=2 (1+\bar\mu_j)^3+(1+\bar\mu_j)^{2}\left[-2+\frac{3}{2}\bar\mu_j-\frac{1}{2}\left(\frac{\Delta}{\Delta-1}\right)\right]
+2\bar\mu_j\left\{3+\left(\frac{\Delta-1}{\Delta}\right) \left[-8(1+\bar\mu_j)+3\bar\mu_j+4\right]\right\}$, and \\
$H_4(\Delta)=2(1+\bar\mu_j)^2+\frac{1}{2}(1+\bar\mu_j)\left[\frac{1}{2}\left(\frac{\Delta}{\Delta-1}\right)+3\bar\mu_j\right]-12\bar\mu_j\left(\frac{\Delta}{\Delta-1}\right)$. \\
The last term in Eq.(\ref{GF2}) accounts for the disorder corrections to the EoS.
For $\Delta\rightarrow\infty$ (or $g_{12} \rightarrow 0$, equivalently), one has $h_j(\infty)=3/4$ (see also Fig.\ref{dis11}) and thus, the EoS reduces to that of the single component BEC 
namely $\mu = g n(1+12\pi R' \sqrt{n a^3/\pi})$, found in Refs.\cite{Falco,Gior,Yuk} using the Huang-Meng-Bogoliubov theory.

Figure \ref{dis11} (a) depicts that the functions $h_j$ grow with $a _{12}$ and diverge at $a_{12} \rightarrow \sqrt{a_1a_2}$
results in an enhancement of the total chemical potential. 
In this case, the quantum fluctuations arising from interactions are viewed as being predominated by disorder effects.

Moreover, we see from Fig.\ref{dis11} (b) that the disorder functions $h_j$ behave differently with the interactions ratio $a_2/a_1$.
Both functions diverge for $a_2/a_1\rightarrow 0$ and match at $a_2/a_1=5$. 
The chemical potential associated with the first component $\mu_1$ enhances  when $h_1$ rises, while $\mu_2$ decays for lowering $h_2$.
This reveals that the competition of the intraspecies interactions and the disorder potential may perceptibly alter the behavior of the  EoS of the whole mixture.

\section{Miscibility conditions} \label{MC}

We now discuss a possible energetic instability, associated with the presence of the disorder and the occurence of miscible-immiscible phase transition.
For a homogeneous mixture to be stable, the following conditions should be fulfilled \cite{PS}:
\begin{subequations}\label{Stable}
\begin{align}
 &\frac{\partial\mu_j}{\partial n_j} > 0, \\
&  \left(\frac{\partial\mu_j}{\partial n_j}\right) \left(\frac{\partial\mu_{\overline j}}{\partial n_{\overline j}}\right) > \left(\frac{\partial\mu_j}{\partial n_{\overline j}}\right)^2.
\end{align}
\end{subequations}
These conditions are derived from the variation of the energy with respect to the densities.
For the EoS (\ref{GF2}), we obtain 
\begin{equation} \label {comp}
 \frac{\partial\mu_j}{\partial n_j}= g_j \left [  1+ 8\pi R'_j \sqrt{\frac{ n_j a_j^3}{\pi}} \left(h_j+2n_j\frac{\partial h_j (\Delta)}{\partial n_j}\right )\right].
\end{equation}
The second term in the r.h.s of Eq.(\ref{comp}) constitutes the disorder corrections to the inverse compressibility $\kappa_j^{-1}= n_j^2\partial \mu_j/\partial n_j$.

\begin{figure}[htb] 
\includegraphics[scale=0.8]{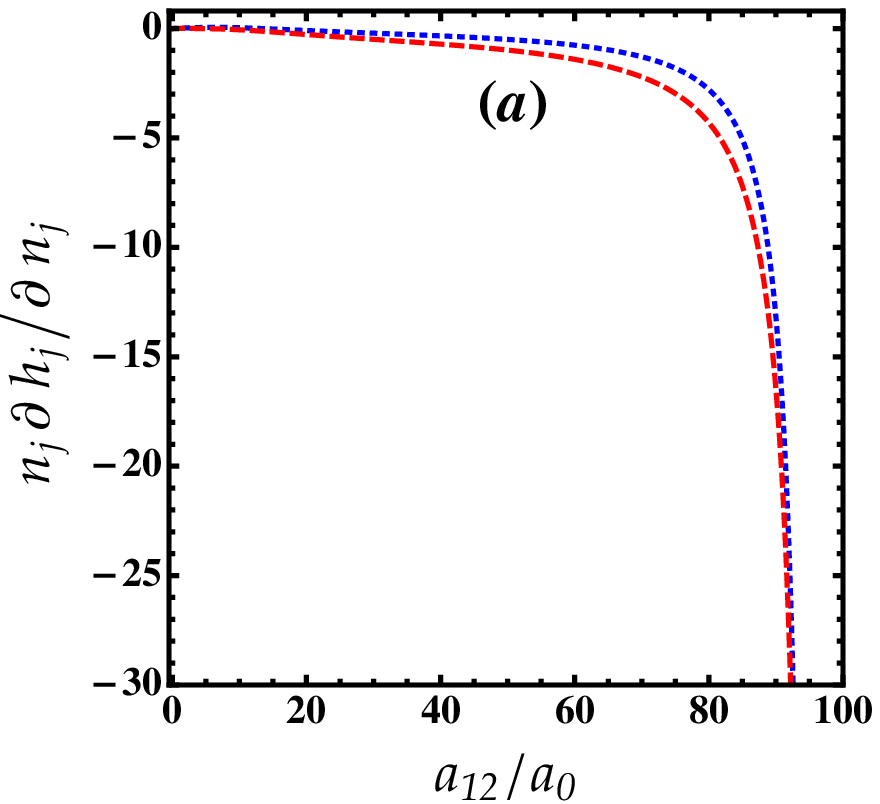}
\includegraphics[scale=0.8]{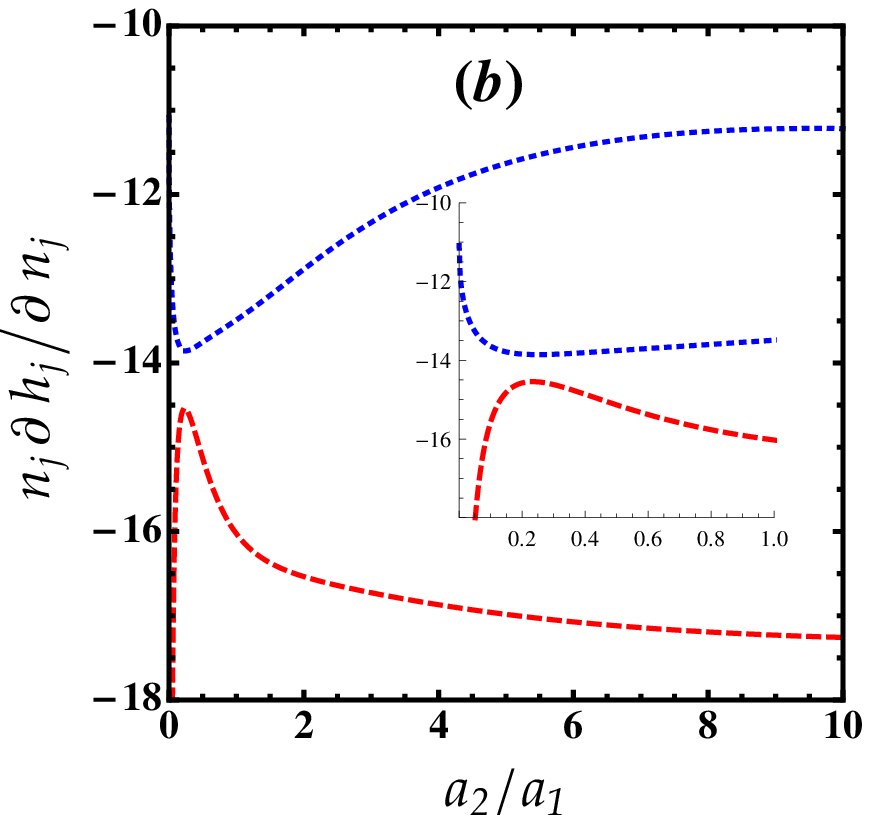}
\caption {(Color online) (a) Behavior of the disorder functions $\partial h_j/\partial n_j$ in units of $n_j$ as a function of $a_{12}$  for ${}^{87}$Rb-${}^{87}$Rb mixture. 
(b) Behavior of the disorder functions $\partial h_j/\partial n_j$ in units of $n_j$ as a function of the ratio $a_2/a_1$ for $a_{12}=90 a_0$.
Blue dotted lines: $n_1 \partial h_1/\partial n_1$. Red dashed lines: $n_2 \partial h_2/\partial n_2$.}
\label{dis111}
\end{figure}

Figure \ref{dis111} (a) shows that the disorder functions $n_j \partial h_j /\partial n_j$ possess identical behavior 
over almost the entire range of the interspecies interactions. 
They vanish for $a_{12}=0$ where the two components are spatially separated and remain
negligibly small in the domain $0 \leq a_{12}/a_0 \lesssim 65 $, indicating that the disorder effect is marginally relevant in this regime.
For $a_{12}/a_0 \gtrsim  65$, $n_j \partial h_j /\partial n_j$ decrease and display a negative divergence at $a_{12} \rightarrow \sqrt{a_1a_2}$, 
leading to appreciably reduce the compressibility of the system.

We observe from  Fig.\ref{dis111} (b) that the disorder functions $n_1 \partial h_1 /\partial n_1$ and $n_2 \partial h_2 /\partial n_2$ vary in the opposite way with the ratio $a_2/a_1$.
They diverge for $a_2/a_1\rightarrow 0$,  and have minimum/maximum at $a_2/a_1 \simeq 0.2$, where the second component is extremely dilute compared to the first component, 
then they increase/decrease for $a_2/a_1 > 0.2$ (see the inset of Fig.\ref{dis111} (b)). 
This peculiar behavior can be attributed to the competition between the repulsive interactions, the miscibility and the disorder.
The functions $\partial h_j /\partial n_j$ are negative in the whole range of interactions.

The stability conditions (\ref{Stable}) turn out to be given as
\begin{subequations}\label{Stabl}
\begin{align}
 &g_j \left [ 1+ 8\pi R'_j \sqrt{\frac{ n_j a_j^3}{\pi}} \left(h_j+2n_j\frac{\partial h_j (\Delta)}{\partial n_j}\right )\right]> 0,  \label{Stable1} \\
\text{and} \nonumber\\
&   \Delta \left [ 1+ 8\pi R'_j \sqrt{\frac{ n_j a_j^3}{\pi}} \left(h_j+2n_j\frac{\partial h_j (\Delta)}{\partial n_j}\right )\right] \nonumber \\
&\times \left [  1+ 8\pi R'_{\overline j} \sqrt{\frac{ n_{\overline j} a_{\overline j}^3}{\pi}} \left(h_{\overline j}+2n_{\overline j}\frac{\partial h_{\overline j} (\Delta)} 
{\partial n_{\overline j}}\right )\right] \nonumber \\
&>\left(1+16\pi R'_j\sqrt{\frac{n_ja_j^{3}}{\pi}}\frac{ a_j} {a_{12}} n_j\frac{ \partial h_j (\Delta)}{\partial n_{\overline j}} \right)^2. \label{Stable2}
\end{align}
\end{subequations}
Expressions (\ref{Stabl}) clearly show that the miscibility condition for a mixture of two interacting BEC is significantly affected by the disorder potential. 
This gives rise to a phase transition to an immiscible phase even though the cleaned mixture is miscible. 
For relatively large disorder strength, the mixture may drive a transition to an immiscible phase with complete spatial separation between the two BEC.
For $R'_j=0$, the conditions (\ref{Stabl}) reduce to those of the cleaned binary BEC mentioned above. 

The critical disorder strength above which a quantum miscible-immiscible phase transition occurs can be directly determined from (\ref{Stable2}) as
\begin{equation} \label {DSrg}
 R'^c_j= \frac{-A_j-\sqrt{A_j^2-4B_j (\Delta-1)/\Delta }}{16 \pi \sqrt{n_ja_j^3/\pi} B_j},
\end{equation} 
where $A_j=(h_j+2n_j \partial h_j/\partial n_j) + \sqrt{g_j n_j/g_{\overline j} n_{\overline j}} (h_{\overline j}+2n_{\overline j} \partial h_{\overline j}/\partial n_{\overline j})
-4\Delta^{-1} (n_j g_j/g_{12})  (\partial h_j / \partial n_{\overline j})$, and 
$B_j=(h_j+2n_j \partial h_j/\partial n_j)  (h_{\overline j}+2n_{\overline j} \partial h_{\overline j}/\partial n_{\overline j})  \sqrt{g_j n_j/g_{\overline j} n_{\overline j}}
-4 \Delta^{-1} (n_j g_j/g_{12})^2  (\partial h_j / \partial n_{\overline j})^2$ with $R_{\overline j}'= R_j' (g_j ^2n_j/g_{\overline j}^2 n_{\overline j})$. 
In the case of ${}^{87}$Rb-${}^{87}$Rb mixture with parameters : $a_1= 100.4 a_0$, $n_1=1.5\times 10^{21}$ m$^{-3}$ and $a_2= 95.44 a_0$, $n_2=10^{21}$ m$^{-3}$, 
and $a_{12}=90 a_0$, the miscible-immiscible phase transition arises for disorder strengths $R'^c_1=0.7$ and $R'^c_2=1.16$.

\section{Superfluid fraction} \label{SF}

Let us consider a Bose mixture superfluid moving with velocity ${\bf v_{sj}}= \hbar {\bf k_{sj}}/m$, where ${\bf k_{sj}}$ is a wavevector corresponding to the velocity of superfluid,
subjected to a moving  weak random potential with the velocity ${\bf v_u}= \hbar {\bf k_u}/m$, where $ {\bf k_u}$ is a wavevector corresponding to the velocity of disorder.
At finite temperatures, the Bose fluid is separated into a superfluid density $n_{sj}$ and a normal density $n_{nj}$ that moves with
the disorder component $n_{Rj}$.
Then the coupled time-dependent GP equations read
\begin{align}\label{TDGPE}
i\hbar\frac{\partial \Phi_{j}( {\bf r},t)} {\partial t}&=\bigg(-\frac{\hbar^{2}}{2m}\nabla^2+U({\bf r}-{\bf v_u}t) \\
&+g_{j}|\Phi_{j}({\bf r},t)|^{2} +g_{12}\vert\Phi_{\overline j} ({\bf r},t)\vert^{2}\bigg) \Phi_j ({\bf r},t). \nonumber
\end{align}
We treat the solution of  Eq.(\ref{TDGPE})  perturbatively  by introducing the function  
\begin{align}\label{SSExps}
\Phi_j  ({\bf r},t)&=\big[\Phi_j^{(0)}+\Phi_j^{(1)} ({\bf r},t)+\Phi_j^{(2)} ({\bf r},t)+\cdots \big] \\
&\times e^{i\bf k_{sj} . r} e^{-\frac{i}{\hbar} \big(\frac{\hbar^2 {\bf k}^2_{sj}} {2m}+\mu_{j} \big)t}, \nonumber
\end{align}
which corresponds to the clean-case solution \cite{Nik,Boudj4,Gior}.
After inserting  the expansion (\ref{SSExps}) into Eq.(\ref{TDGPE}), and using the transformation ${\bf r'}={\bf r}+{\bf v_u} t$, one obtains
\begin{align}\label{TDGPE1}
\bigg(-\frac{\hbar^2}{2m}\nabla^2- i\frac{\hbar^2}{m} {\bf K}_j.{\bf \nabla} +U({\bf r'})-\mu_j+g_{j}\vert\Phi_j ({\bf r'})\vert^{2} \\
+g_{12}\vert\Phi_{\overline j} ({\bf r'})\vert^2\bigg)\Phi_j ({\bf r'})=0, \nonumber
\end{align}
where ${\bf K_j}={\bf k_{sj}} -{\bf k_u}$. \\
In the two-fluid model, the total momentum ${\bf P (r)}$ of the moving system is defined as: 
\begin{equation}\label{Mom}
{\bf P}_j=-i\hbar\langle \Phi_{j}\vert i {\bf k}_{sj}+\nabla\vert\Phi_{j}\rangle =\hbar {\bf k}_{sj} n_j-i\hbar\langle\Phi_{j}^{*}\nabla\Phi_{j}\rangle.
\end{equation}
We neglect higher than linear terms in ${\bf k}_{sj}$ and keeping in mind that in zeroth order ${\bf P}_j$ does not depend on ${\bf k}_{sj}$. 
This yields
\begin{equation}\label{Mom1}
{\bf P}_j=\hbar {\bf k}_{sj} n_j -i\hbar\langle\Phi^{*(1)}_j\nabla\Phi_j^{(1)} \rangle+ \cdots,
\end{equation}
where the first-order correction to the wavefunction is given in Fourier space by
\begin{widetext} 
\begin{equation}\label{E36}
\Phi_j^{(1)}({\bf k})=\frac{-U({\bf k})\Phi_{j}^{(0)}(E_k-\frac{\hbar^{2}}{m}{\bf k .K_j})\left[-E_k^2-2E_k\Phi_{\overline j}^{(0)2}(g_{\overline j}-g_{12})+(\frac{\hbar^{2}}{m}
 {\bf k.K_{\overline j}})^{2}\right]}
{4\Phi_{\overline j}^{(0)2}\Phi_j^{(0)2} g_{12}^2 E_k^2-\left(E_k^2+2E_k g_{\overline j} \Phi_{\overline j}^{(0)2}-(\frac{\hbar^{2}}{m} {\bf k. K_{\overline j}})^2 \right)
\left(E_k^{2}+2E_k g_j\Phi_j^{(0)2}-(\frac{\hbar^{2}}{m} {\bf k. K_j})^{2}\right)}.
\end{equation}
\end{widetext} 
 For small $K_j$, the normal density reads
\begin{equation}\label{NS} 
n_{nj}=n_j-\frac{1}{\hbar}\frac{\partial P_{j}}{\partial K_{j}}\Big\vert_{K_j=0}.
\end{equation}
In the case of delta-correlated random potential (\ref{delt}), we get for the normal fraction
\begin{equation}\label{NS1} 
n_{nj}=\frac{16 \pi}{3} R_j' \sqrt{\frac{ n_j a_j^3}{\pi}}  f_j(\Delta)=\frac{4}{3} n_{Rj}.
\end{equation}
We see that Eq.(\ref{NS1}) well recovers the result of Huang-Meng for a single component BEC with contact interaction \cite{HM}. 
The fact that $n_{nj}$ is larger than $n_{Rj}$ is due to the localization of bosons in the respective minima of the random potential which leads to reduction of the superfluid density.  
Obviously, the interplay of the disorder potential, interspecies interaction and the ratio of intraspecies interactions may strongly affect the superfluid fraction $n_{sj}=1- (4/3)n_{Rj}$.

\section{Conclusions} \label{Conc}

We investigated the impact of a weak disorder potential with a delta-correlated function of a homogeneous binary BEC at zero temperature. 
Within the realm of the perturbative theory, we derived analytical expressions for the physical quantities of interest
such as the condensate depletion due to the disorder, the EoS, the compressibility, and the superfluid density in terms of density, strength of disorder and the miscibility parameter.  
Our results revealed that the intriguing interplay of the disorder and intra- and interspecies coupling may strongly influence both the quantum fluctuations and the superfluidity
yielding a variety of interesting situations for relevant experimental parameters. 
In particular we found either both species are localized, or only one species is localized and the second species remains extended.
We showed in addition that the localization of one component does not necessarily trigger the localization of the other species.
Interestingly, we found that the disorder potential leads to a dramatic phase separation between the two species,  
changing the miscibility criterion of the mixture.
We expect that the introduction of the Lee-Huang-Yang (LHY) corrections that stem from quantum fluctuations in the EoS \cite{Gior1} may stabilize the miscible state analogously 
to the quantum mechanical stabilization of the droplet phase \cite{Petrov}. 
The same scenario takes place in a disordered dipolar BEC with the LHY quantum corrections \cite{Boudj4}. 
Furthermore, as in the disordered single BEC, the disorder corrections to the normal part of each Bose fluid have been found to be greater than 
the disorder condensate depletion in each species because the bosons scattered by the disorder environment provide randomly distributed obstacles for the motion of the superfluid.
The results obtained by Huang and Meng and the perturbation theory in a single BEC for the fluctuations of the condensate 
and of the superfluid density due to the disorder have been well-recovered. 

Strictly speaking, in the regime of a strong disorder, each component fragments into a number of low-energy, localized single-particle states
with no gauge symmetry breaking forming the so-called Bose glass phase. 
The exploration of such a regime would need either a non-perturbative approach or Quantum Monte Carlo simulations.
We believe that the findings of this work add extra richness to the diversity of disordered ultracold atoms. 
They open up a new avenue for controlling phase separation of Bose-Bose mixtures.
Finally, an important extension of this work would be to analyze effects of a weak disorder in a mixture droplet state.

\section*{Acknowledgments}
We acknowledge  Axel Pelster and Laurent Sanchez-Palencia for fruitful discussions and useful comments about the paper.

\end{document}